\documentclass[twocolumn]{aastex631}

\usepackage{CJK}
\usepackage{placeins}
\usepackage{amsmath}
\usepackage{cleveref}
\usepackage{soul}

\shorttitle{Cosmology from one galaxy in a void}
\shortauthors{B. Y. Wang \& A. Pisani}

\graphicspath{{./}{./Figures/}}

\begin{document}
\begin{CJK*}{UTF8}{bsmi}

\title{Cosmology from one galaxy in a void?}

\correspondingauthor{Bonny Y. Wang}
\email{yuew7@andrew.cmu.edu}

\author[0000-0001-7168-8517]{Bonny Y. Wang (汪玥)}
\affiliation{Carnegie Mellon University, 5000 Forbes Ave, Pittsburgh, PA 15213}
\affiliation{Center for Computational Astrophysics, Flatiron Institute, 162 5th Avenue, New York, NY 10010 USA}

\author[0000-0002-6146-4437]{Alice Pisani}
\affiliation{Center for Computational Astrophysics, Flatiron Institute, 162 5th Avenue, New York, NY 10010 USA}
\affiliation{Aix-Marseille University, CNRS/IN2P3, CPPM, 163 Av. de Luminy, 13009, Marseille, France}
\affiliation{The Cooper Union for the Advancement of Science and Art, 30 Cooper Sq, New York, NY 10003 USA}
\affiliation{Department of Astrophysical Sciences, Princeton University, 4 Ivy Lane, Princeton, NJ 08544 USA}

\begin{abstract}

Understanding galaxy properties may be the key to unlocking some of the most intriguing mysteries of modern cosmology. Recent work relied on machine learning to extract cosmological constraints on $\Omega_\mathrm{m}$ using only \textit{one} galaxy. But if this is true, how should we select \textit{the} galaxy to use for cosmology inference? In this paper, we consider selecting a galaxy that lies in cosmic voids, the underdense regions of the cosmic web, and compare the constraints obtained with the ones obtained when randomly selecting a galaxy in the whole sample. We use the IllustrisTNG galaxy catalog from the CAMELS project and the {\tt\string VIDE} void finder to identify galaxies inside voids. We show that void galaxies provide stronger constraints on $\Omega_\mathrm{m}$ compared to randomly selected galaxies. This result suggests that the distinctive characteristics of void galaxies may provide a cleaner and more effective environment for extracting cosmological information.
\end{abstract}


\section{Introduction} \label{sec:intro}
One of the primary goals of cosmology is to constrain the parameters of the cosmological model. The matter density parameter $\Omega_\mathrm{m}$, representing the fractional energy density of matter in the Universe, is a crucial parameter to determine in the quest to understand the composition and evolution of our Universe. The current framework for the standard model of cosmology is confirmed by a wide range of cosmological observations, including (but not limited to) galaxy clustering statistics \citep[e.g.][]{Eisenstein2005,Vikhlinin2009, Allen2011,DESI2024Cosmo}, cosmic microwave background \citep[e.g.][]{Hu2002, Spergel2003, Spergel2007, Planck2020} and supernovae \citep[e.g.][]{Phillips1993, Perlmutter1997, Branch1998, Riess1998, Perlmutter1999, Tonry2003, Astier2006, Betoule2014}.

Recently, relying on machine learning, \citet{Villaescusa2022} extracted cosmological constraints from the properties of one galaxy. Subsequently, \citet{Chawak2023} showed that constraints can be improved when using multiple galaxies. This method has also been tested in \citet{Echeverri2023} using four different simulations, including IllustrisTNG \citep{Marinacci2018,Naiman2018,Springel2018, Nelson2019}, SIMBA \citep{Dave2019}, Astrid \citep{Ni2022, Bird2022, Ni2023}, and Magneticum \citep{Dolag2015}. The fact that these simulations all have different subgrid physics models and different redshifts further proved the robustness of the methodology. \citet{Hahn2023} also showed that we can extend this study to observations by extracting cosmological information simply relying on the photometry of a few galaxies.

These results embed underlying questions that still need to be answered: How should we select the one galaxy or group of galaxies to optimize the constraints? Are all galaxies equal in terms of the cosmological information they bear? In this paper, we address this question by considering galaxies in the underdense regions of our Universe, cosmic voids.
    
Voids are sensitive probes to extract cosmological information \citep[e.g.][]{Park2007, Lavaux2010, Biswas2010, Lavaux2012, Pisani2015, Hamaus2016, Pisani2019, Verza2019,Stopyra2021,Kreisch2021, Contarini2022b, Contarini2022, Pelliciari2022, Schuster2023, Wang2023}: for instance the void-galaxy cross-correlation function and the void size function already provide tight constraints on $\Omega_\mathrm{m}$ \citep{Hamaus2020,Contarini2022}. Since voids are goldmines of cosmological information \citep{Pisani2019, Moresco2022}, it is interesting to ask whether galaxies inside voids bear a stronger constraining power from a cosmology perspective. 

Void galaxies are generally expected to be low-mass, blue, and with high star formation rates (\citet{Grogin1999, Rojas2004, Rojas2005, Hoyle2005, Park2009, Krechel2011, Florez2021}. Some studies also suggest void galaxies may have lower metallicities \citep{Pustilnik2003, Pustilnik2005, Pustilnik2011, Rosas2022}. However, whether void galaxies differ from comparable objects in denser regions remains controversial (see e.g. \citet{Kreckel2015, Penny2015,Habouzit2020}). The controversy can also be caused by a lack of a robust framework for studying void substructures and dynamics \citep{Colberg2008}. Still, due to their isolated nature, it is reasonable to expect that void galaxies---less influenced by external disturbances such as stripping and harassment---could have a stronger connection to cosmology and the physics of the early universe. In addition, differences have also been investigated specifically in the IllustrisTNG simulations recently: \citet{Rodriguez-Medrano2023} and \citet{Curtis2024} suggested that void galaxies have significantly different properties compared to non-void galaxies, which may impact the link to cosmological information. 

In this work, we investigate whether individual void galaxies contain more pristine information about cosmology than galaxies residing in other environments. Section \ref{sec:methods} introduces the simulations and datasets, as well as the method we used to select the void and non-void galaxies. Section \ref{sec:results} illustrates our predictions for $\Omega_\mathrm{m}$ when using void and non-void galaxies. Finally, Section \ref{sec:conc} concludes and discusses future prospects.

\section{Methods} \label{sec:methods}

\subsection{Simulations and Datasets}
In this work, we used the dataset from the Cosmology and Astrophysics with MachinE Learning Simulations (CAMELS) project, which includes more than 10,000 numerical simulations with different cosmological and astrophysical models \citep{Villaescusa2021}. Specifically, we use the galaxy catalogs from the LH set of the IllustrisTNG suite. The IllustrisTNG simulations are run with the AREPO code \citep{Weinberger2020}.
This suite contains 1000 simulations of $(25 h^{-1}Mpc)^3$ volume each and, while the values of $\Omega_{\rm b}=0.049$, $h=0.6711$, $n_s=0.9624$, $\sum m_\nu=0.0$ eV, $w=-1$ are fixed, the values of other cosmological and astrophysical parameters are sampled from a latin hypercube (LH) with different initial random seeds.
In particular, the matter density parameter $\Omega_\mathrm{m}$ is varied within the range of 0.1 to 0.5. Notably, each simulation also varies four astrophysical parameters: two for supernovae ($A_{\rm SN1}$ and $A_{\rm SN2}$) and two for AGN ($A_{\rm AGN1}$ and $A_{\rm AGN2}$). Therefore, when inferring $\Omega_\mathrm{m}$, we implicitly marginalize over baryonic uncertainties.

We follow the procedure described in \citet{Villaescusa2022}, and consider the following 17 galaxy properties:
\begin{enumerate}
\item $M_{\rm g}$: The mass of gas in the galaxy, including contributions from the circumgalactic medium.
\item $M_{\rm BH}$: The mass of the galaxy's black hole.
\item $M_*$: The stellar mass of the galaxy.
\item $M_{\rm t}$: The total mass of the subhalo hosting the galaxy, comprising dark matter, gas, stars, and black holes.
\item $V_{\rm max}$: The maximum circular velocity of the galaxy's subhalo, defined as $V_{\rm max}=\max(\sqrt{GM(<R)/R}$).
\item $\sigma_v$: The velocity dispersion of all particles within the galaxy's subhalo.
\item $Z_{\rm g}$: The mass-weighted metallicity of the galaxy's gas.
\item $Z_*$: The mass-weighted metallicity of the galaxy's stars.
\item ${\rm SFR}$: The star formation rate of the galaxy.
\item $J$: The magnitude of the spin vector of the galaxy's subhalo.
\item $V$: The magnitude of the peculiar velocity of the galaxy's subhalo.
\item $R_*$: The radius that contains half of the galaxy's stellar mass.
\item $R_{\rm t}$: The radius that contains half of the total mass of the galaxy's subhalo.
\item $R_{\rm max}$: The radius at which $\sqrt{GM(<R_{\rm max})/R_{\rm max}}=V_{\rm max}$.
\item ${\rm U}$: The galaxy's magnitude in the U band.
\item ${\rm K}$: The galaxy's magnitude in the K band.
\item ${\rm g}$: The galaxy's magnitude in the g band.
\end{enumerate}
In addition, we also consider a galaxy as a subhalo that contains no less than 20 particles as stated in \citet{Villaescusa2022}. These galaxies have masses ranging from $\sim 10^8 \mathrm{M}_{\odot}$ to $\sim 10^{14} \mathrm{M}_{\odot}$ with $\sim 80\%$ of galaxies under $\sim 10^{11} \mathrm{M}_{\odot}$ and $\sim 25\%$ of galaxies under $\sim 10^{10} \mathrm{M}_{\odot}$.
\subsection{Predicting $\Omega_\mathrm{m}$ with neural networks using one galaxy}
We follow \citet{Villaescusa2022} and use fully connected layers with hyperparameters optimized by \texttt{optuna} \citep{Akiba2019} to perform likelihood-free inference on $\Omega_\mathrm{m}$ using one galaxy. In order to fully utilize the dataset and train the networks to be adaptable to as many galaxies as possible, we still use all galaxies in the simulation to train our networks. Therefore, the training process is the same as \citet{Villaescusa2022} and as a first test we have reproduced their results by using the public code available at \url{https://github.com/franciscovillaescusa/Cosmo1gal}. We use similar accuracy and precision metrics as in \citet{Echeverri2023} to compare between void and non-void galaxies on predicting $\Omega_\mathrm{m}$. In particular, we rely on the root mean square error (RMSE), the mean relative error $\epsilon$ and the coefficient of determination $R^2$, defined as
\begin{equation}
    \label{eq:acc}
    R^2 = 1-\frac{\langle(\theta - \mu)^2\rangle}{\langle(\theta - \Bar{\theta})^2\rangle},
\end{equation}
\begin{equation}
    \label{eq:acc}
    \text{RMSE} = \sqrt{\langle(\theta - \mu)^2\rangle},
\end{equation}
\begin{equation}
    \epsilon = \left\langle\frac{\sigma}{\mu}\right\rangle,
\end{equation}
where the $\mu$, $\sigma$ are the predicted posterior mean and standard deviation of $\Omega_\mathrm{m}$ from the networks, $\theta$ is the true value of $\Omega_\mathrm{m}$, and $\Bar{\theta}$ is the average value of true values. A lower value for the RMSE indicates that the model is more accurate, and the value of $\epsilon$ is smaller for a more precise model. Finally, a value of $R^2$ closer to 1 indicates a more accurate model.

\subsection{The Void Finder}
\label{sec:voidfinder}
To build a void catalog from the simulations, we use the {\tt\string VIDE} void finder\footnote{\url{https://bitbucket.org/cosmicvoids/vide_public/wiki/Home}} \citep{Sutter2015}. {\tt\string VIDE} is a popular public Voronoi-watershed void finder toolkit based on {\tt\string ZOBOV} \citep{Neyrinck2008}. 
It has been widely used in the field \citep[see e.g.][]{Hamaus2015,Hamaus2016, Hamaus2017,Hamaus2022, Baldi2016,Habouzit2020,Kreisch2021,Contarini2022,Doglass2022}. {\tt\string VIDE} allows to find voids with no assumption on the shape, relying on topological information. 

The void finding procedure has various steps: first, it performs a Voronoi tessellation of the galaxy field. Since regions with fewer (more) galaxies will correspond to larger (smaller) Voronoi cells, this first step provides a local density field estimator through the inverse of the Voronoi cell volume. Once Voronoi cells are available, {\tt\string VIDE} builds the void regions by merging Voronoi cells: it starts with local minima (largest cells) and merges adjacent cells if the density increases, the merging is done through the watershed transform and provides the voids for the analysis. When cells are merged to form voids, it is then possible to define a void center (calculated as the volume weighted barycenter of the Voronoi cells for each void). We note that the center, being defined through the whole void, bears information about the large-scale structure environment of the void. 

It is also important to notice that the definition of voids from {\tt\string VIDE} includes overdense regions constituting the outer rim at the edge of voids (as detailed in \citet[][]{Habouzit2020,Zaidouni2024}). Therefore the collection of Voronoi cells that we use will include both cells from the core region of voids, as well as Voronoi cells on the outer regions, in more overdense environments. The volume of each Voronoi cell is a measure of how isolated the galaxy corresponding to that cell is: the larger the volume, the more isolated the galaxy. 

\begin{figure}
    \centering
    \includegraphics[scale=0.5]{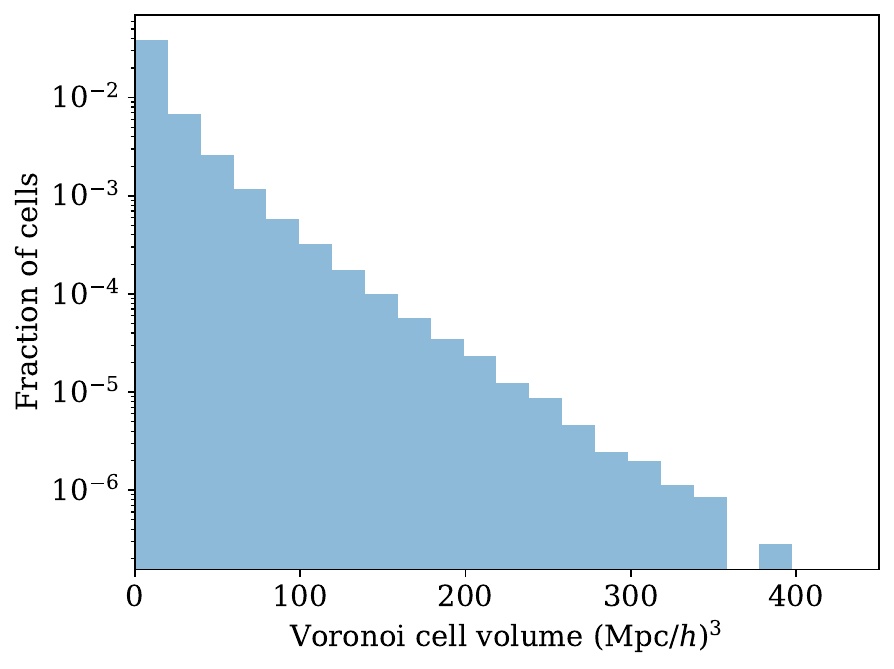}
    \caption{Histogram of Voronoi cell volume in voids for galaxies in the dataset. Galaxies with larger Voronoi cell volume are more isolated.}
    \label{fig:cellDis}
\end{figure}

In this paper, we carefully need to select galaxies based on both their local environment and the large-scale environment. While the void center position (and any selection at a given distance from it) allows to account for the large-scale environment, the Voronoi cell volume guarantees a detailed galaxy selection that also accounts for their local environment. The distribution of the Voronoi cell volume for galaxies in our dataset is shown in Figure \ref{fig:cellDis}. 

A careful selection of void galaxies is particularly relevant here, given the relatively small simulation box of $(25 ~h^{-1}{\rm Mpc})^3$ (compared to the size of cosmic voids, that usually span sizes in the range of a few to a hundred of $~h^{-1}{\rm Mpc}$), and to account for the complex shape of voids. Details of how this selection is performed are provided in the following section.

\subsection{Void Galaxy Selection}
\label{sec:Void Galaxy Selection}

We jointly account for the large-scale structure and the local environment and perform a conservative selection of void galaxies in isolated environments. Additionally, we consider purity cuts to avoid the presence of spurious voids.

First, considering the void catalogs with no cuts, we begin with the set of galaxies belonging to {\tt\string VIDE} voids. Since {\tt\string VIDE} voids will include the overdense shells constituting the outer regions of voids, we perform a first selection of galaxies based on the distance from the void center, considering void galaxies as the ones within 75\% of the void radius. The void radius is defined as $R = \left(\frac{3}{4\pi}V\right)^{1/3}$, where $V$ is the total volume of the Voronoi cells that the void contains. We note that, although this cut is based on a radius selection, the void center definition depends on the whole non-spherical shape of the void. 
           
Second, to fully account for void shape for enhanced robustness, we rely on a criteria that is not based on radius, and utilize the Voronoi cell volume as an additional property to select void galaxies. Since larger Voronoi cell volumes correspond to more isolated galaxies, we consider different threshold values for Voronoi cell volumes to select void galaxies. The comparison of results with more or less stringent selection based on the Voronoi cell volumes is particularly instructive as it shows how results vary when considering more isolated galaxies. 
We show results with the various Voronoi cell volume thresholds in Section \ref{sec:results}. 
In the remainder of the paper, we consider both the above criteria (cuts based on distance from the void center and on Voronoi cell volume) to separate void galaxies from non-void galaxies. 

Finally, we add to the above selection additional purity cuts aimed at an enhanced robustness to the presence of spurious voids \citep{Neyrinck2008,Pisani2015b,Cousinou2019}. These purity cuts aim to reduce the presence of galaxies belonging to spurious voids. Previous work showed that spurious voids are more likely to be voids of low size with a shallow density profile---that is a high density at their core \citep{Cousinou2019}. 
Therefore we exclude voids based on the void core density, defined as the density of the core particle in {\tt\string ZOBOV}-normalized units \citep{Neyrinck2008,Sutter2015}, as well as on the void radius. 
These additional cuts allow us to reduce the impact of spurious voids and provide a cleaner void catalog. We use each of these cuts to filter out $\sim 50\%$ of the voids from the catalogs. In Section \ref{sec:results} we show the results obtained when considering only a purity cut on void core density, and when considering both purity cuts on void core density and radius. We checked that our results are robust to variations in the aforementioned cuts.

\section{Results} 
\label{sec:results}

\begin{figure*}[!htb]
    \centering
    \text{\> \> \> Non-void Galaxies \> \> \> \> \> \>  \> \> \> \> \> \> \> \> \> \> \> \> Random Galaxies  \> \> \> \> \> \>  \> \> \> \> \> \> \> \> \> \> \> \> \> \> Void Galaxies}\par\medskip
    \includegraphics[scale= 0.4]{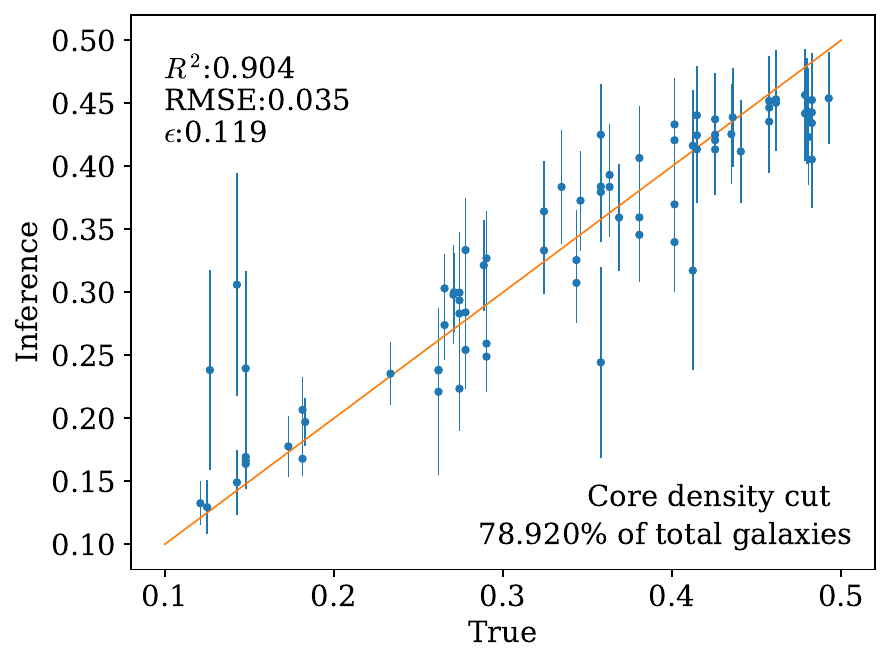}
    \includegraphics[scale= 0.4]{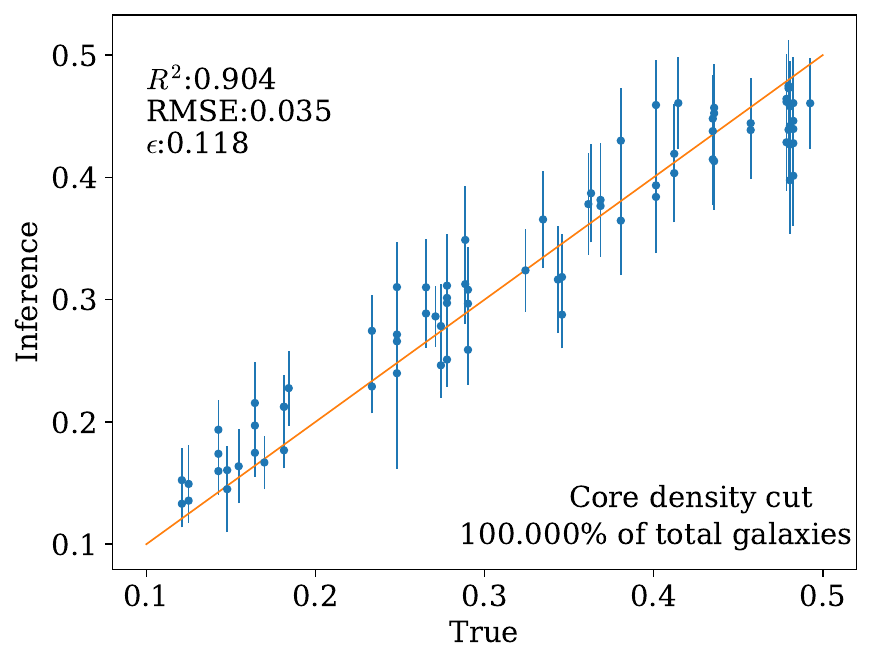}
    \includegraphics[scale= 0.4]{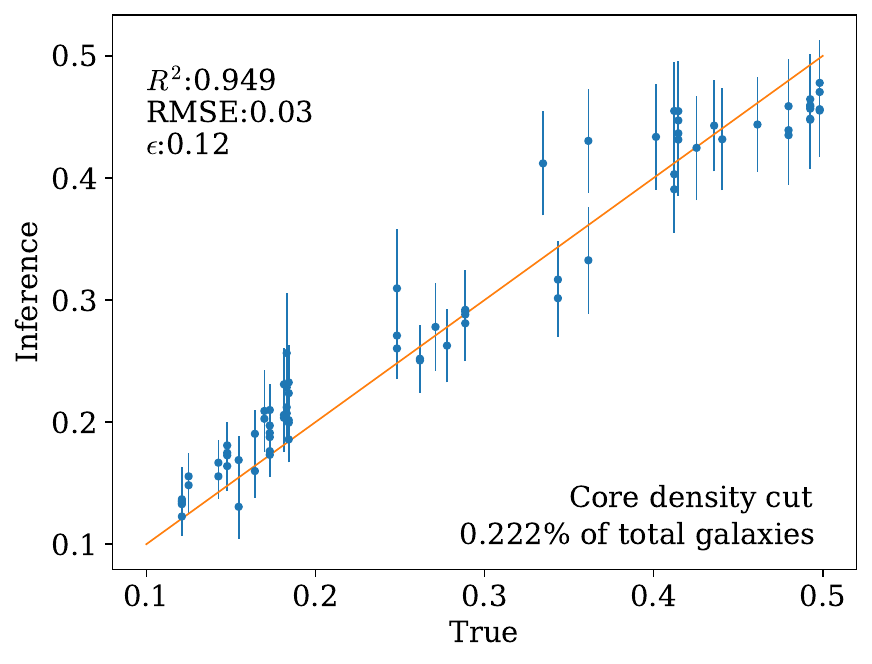}
    \includegraphics[scale= 0.39]{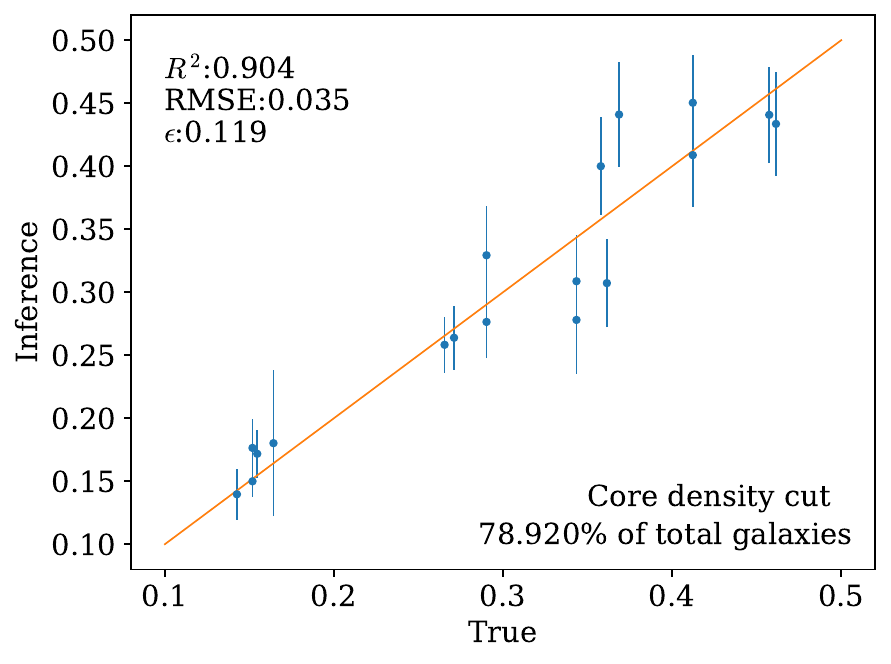}
    \includegraphics[scale= 0.39]{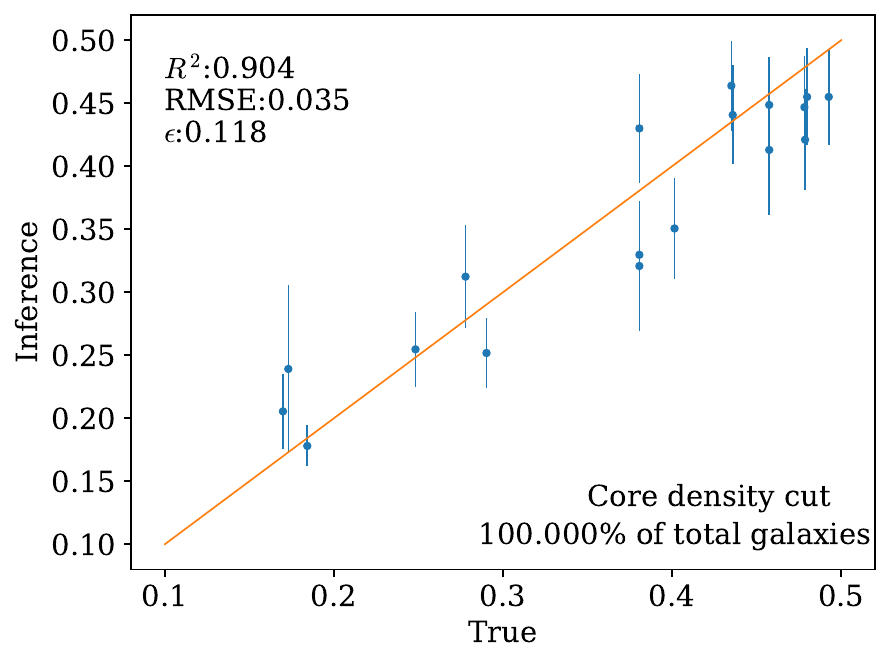}
    \includegraphics[scale= 0.39]{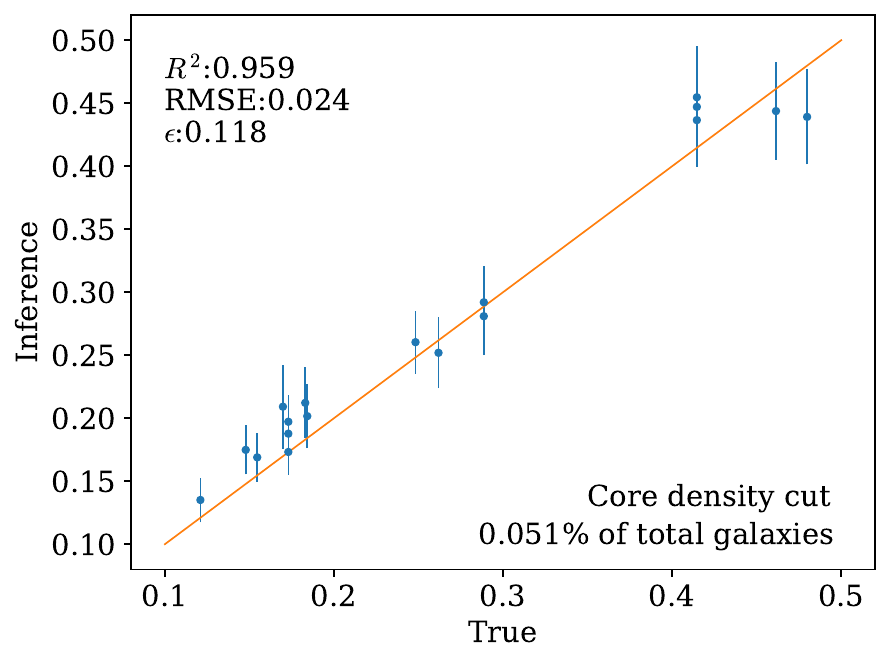}
    
    \caption{$\Omega_\mathrm{m}$ predictions using only one galaxy with fully connected layers. Blue dots and error bars represent the predicted posterior mean and standard deviation of $\Omega_\mathrm{m}$ from the networks. The orange line in the diagonal indicated the perfect match between predicted and true values. The left, middle and right columns show the results when using non-void galaxies, randomly selected galaxies and void galaxies respectively; therefore moving from left to right shows the change when considering the \textit{large-scale environment}. For the right column, the top row shows the results after selecting void galaxies with a cut-off value of Voronoi cell volume at 100 $(\mathrm{Mpc}/h)^3$, while the bottom row shows a more restrictive Voronoi cell volume cut at 150 $(\mathrm{Mpc}/h)^3$, corresponding to more isolated galaxies and accounting for the \textit{local environment}.
    These results correspond to the void catalog considering the purity cut on core density. For better visual comparison, in the case of non-void and randomly selected galaxies we sub-sample the number of points to the same number of data points shown in the plot for void galaxies. The $R^2$ and RMSE values improve for void galaxies (from left to right), and further improve if we consider more isolated galaxies among the void galaxies (from top to bottom, right column).}
    \label{fig:CoreDensCut}
\end{figure*}

\begin{figure*}[!htb]
    \centering
    \text{\> \> \> Non-void Galaxies \> \> \> \> \> \>  \> \> \> \> \> \> \> \> \> \> \> \> Random Galaxies  \> \> \> \> \> \>  \> \> \> \> \> \> \> \> \> \> \> \> \> \> Void Galaxies}\par\medskip
    \includegraphics[scale= 0.39]{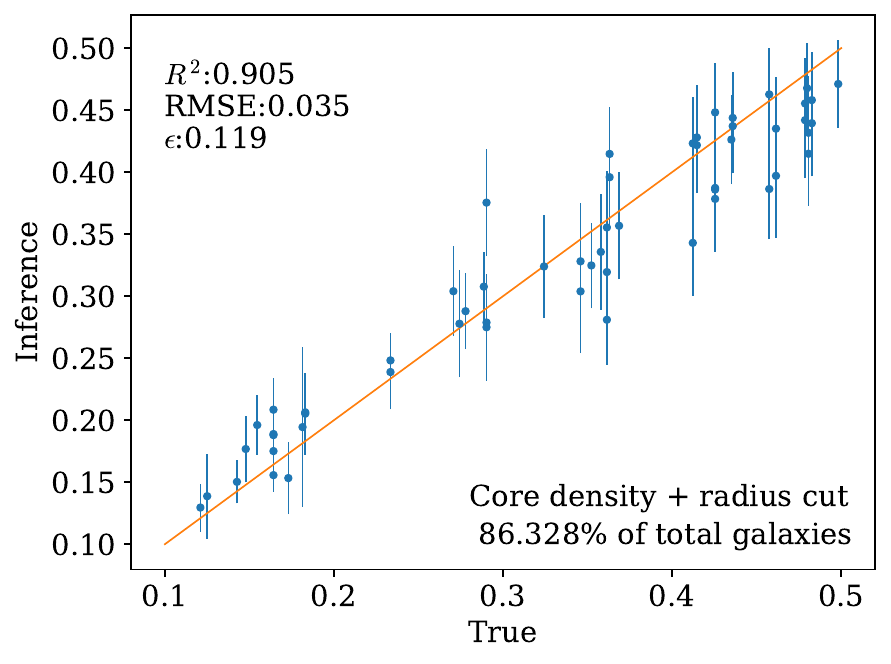}
    \includegraphics[scale= 0.39]{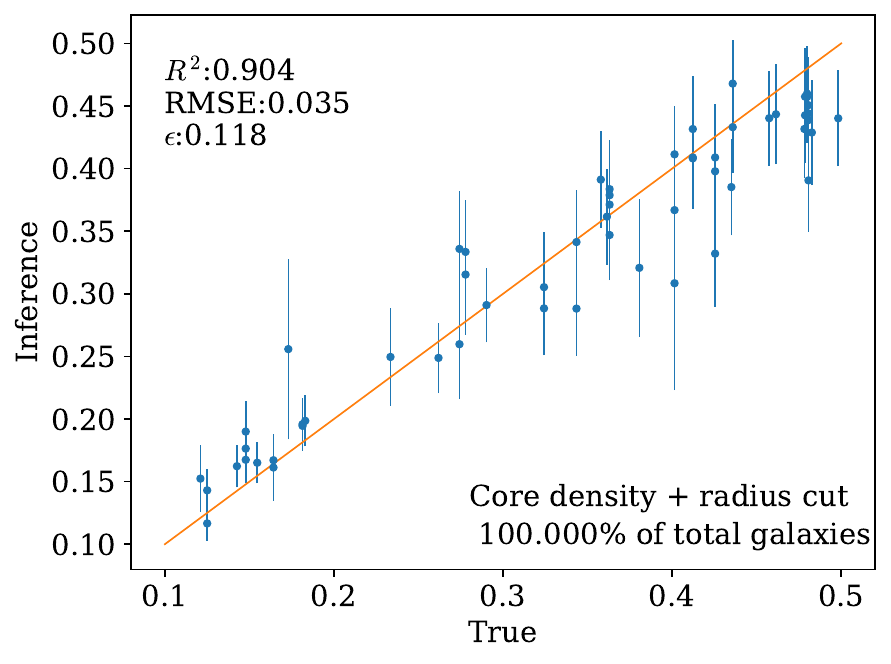}
    \includegraphics[scale= 0.39]{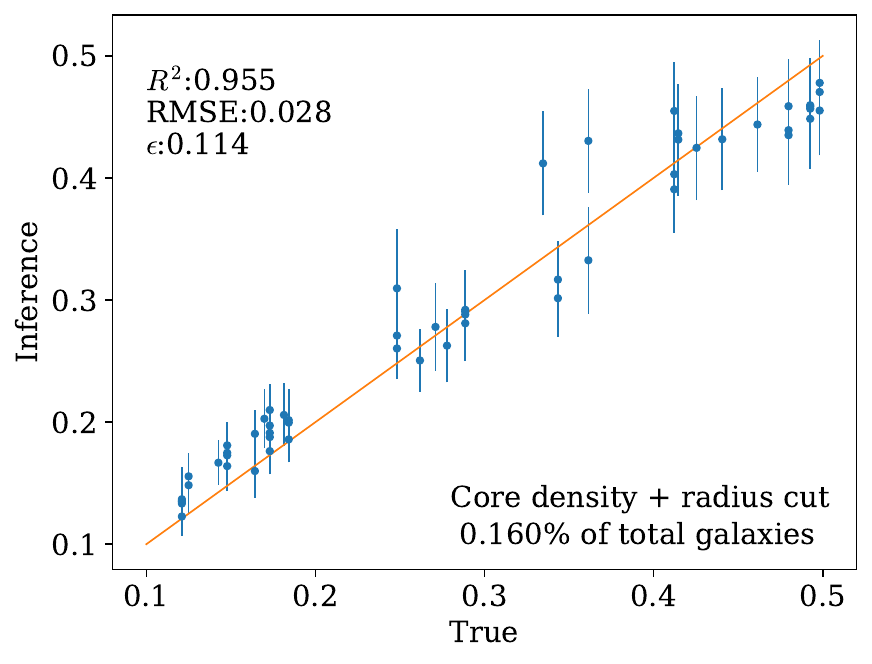}
    \includegraphics[scale= 0.39]{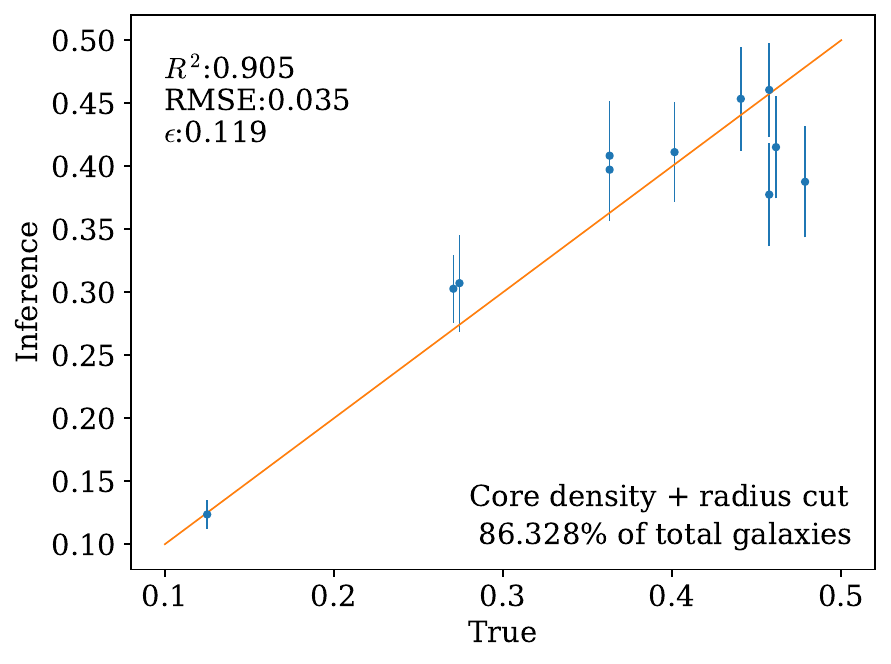}
    \includegraphics[scale= 0.39]{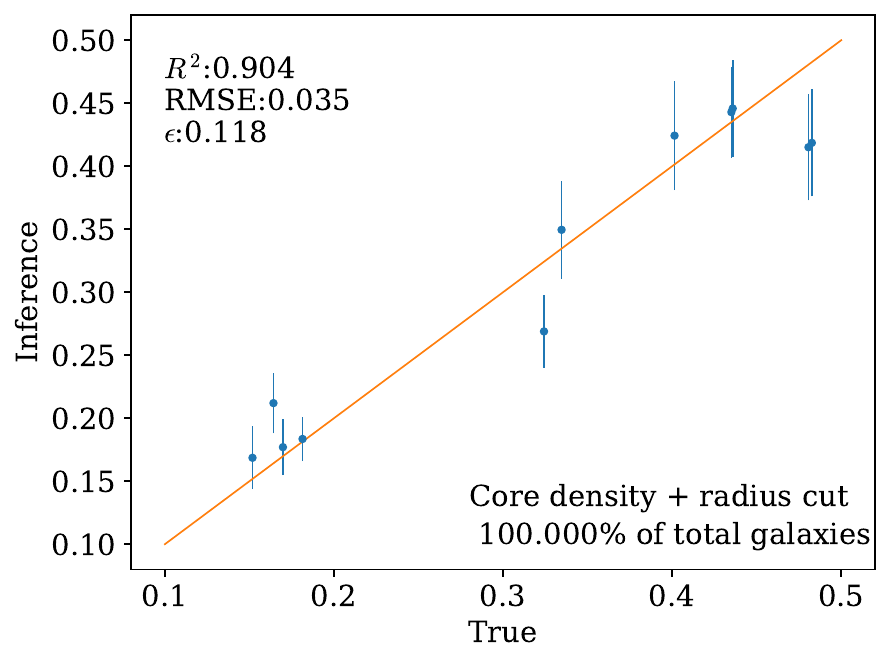}
    \includegraphics[scale= 0.39]{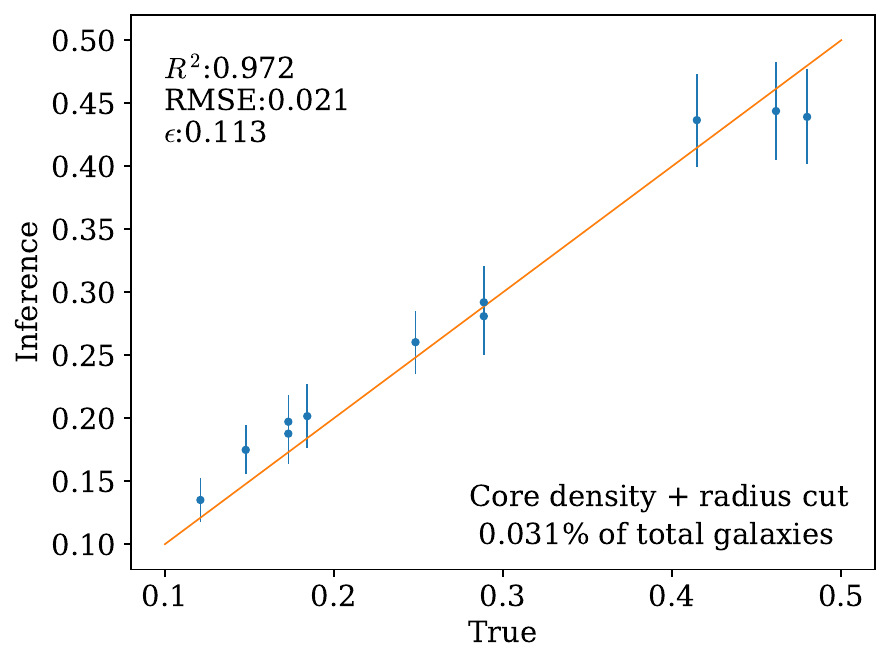}
    \caption{Similar to Figure \ref{fig:CoreDensCut} with the additional radius cut applied for purity to the void catalog. Here, $R^2$, RMSE and $\epsilon$ values all improve for void galaxies (from left to right, accounting for the large-scale environment), and further improve when considering more isolated galaxies among void galaxies, that is considering the local environment (right column, from top to bottom). The strongest improvement is obtained when accounting for selections on both the large-scale and local environments (right column, bottom plot).}
    \label{fig:coreDensityRadiusCut}
\end{figure*}
\begin{figure*}[!htb]
    \centering
    \text{\> \> \> Non-void Galaxies \> \> \> \> \> \>  \> \> \> \> \> \> \> \> \> \> \> \> Random Galaxies  \> \> \> \> \> \>  \> \> \> \> \> \> \> \> \> \> \> \> \> \> Void Galaxies}\par\medskip
    \includegraphics[scale= 0.39]{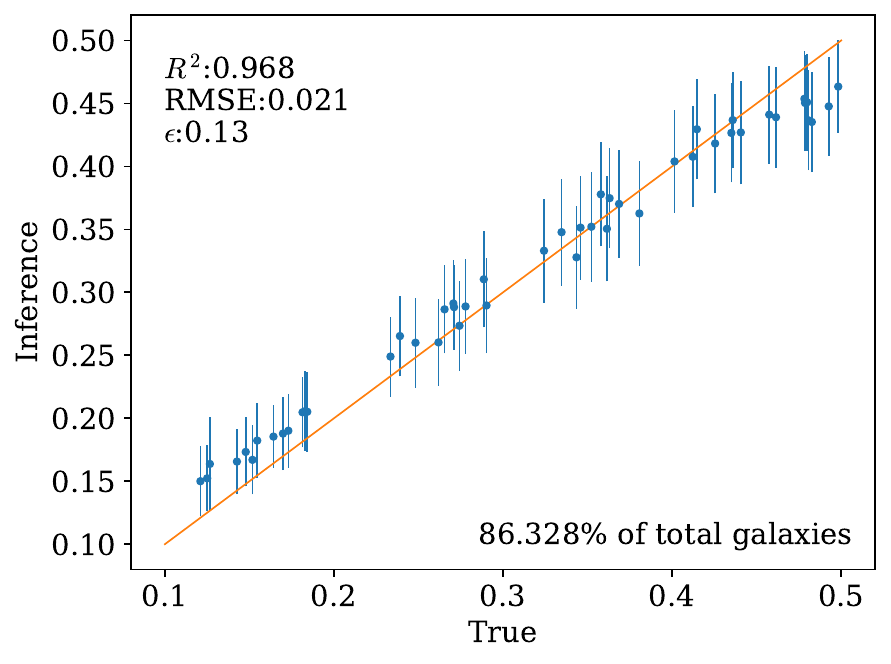}
    \includegraphics[scale= 0.39]{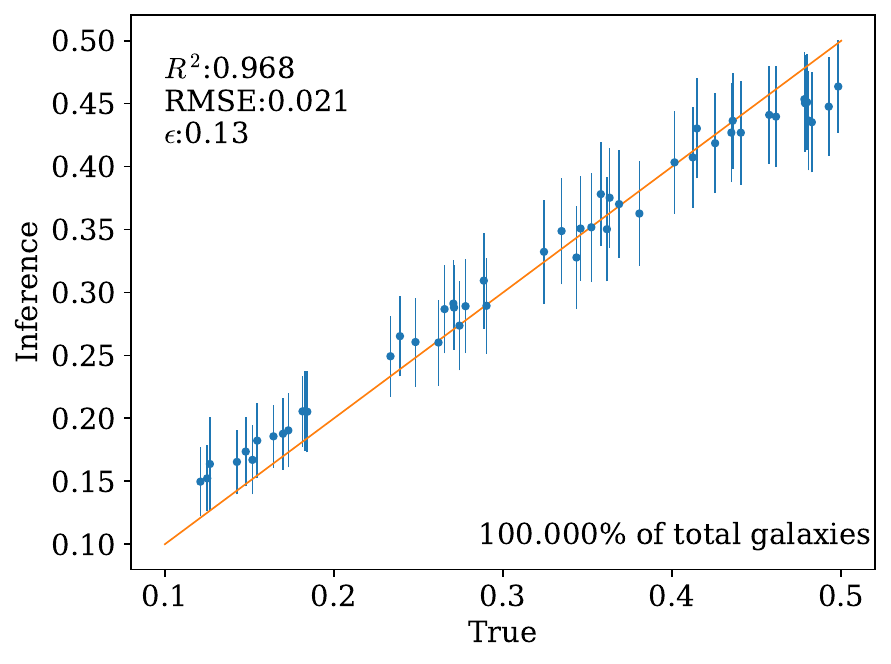}
    \includegraphics[scale= 0.39]{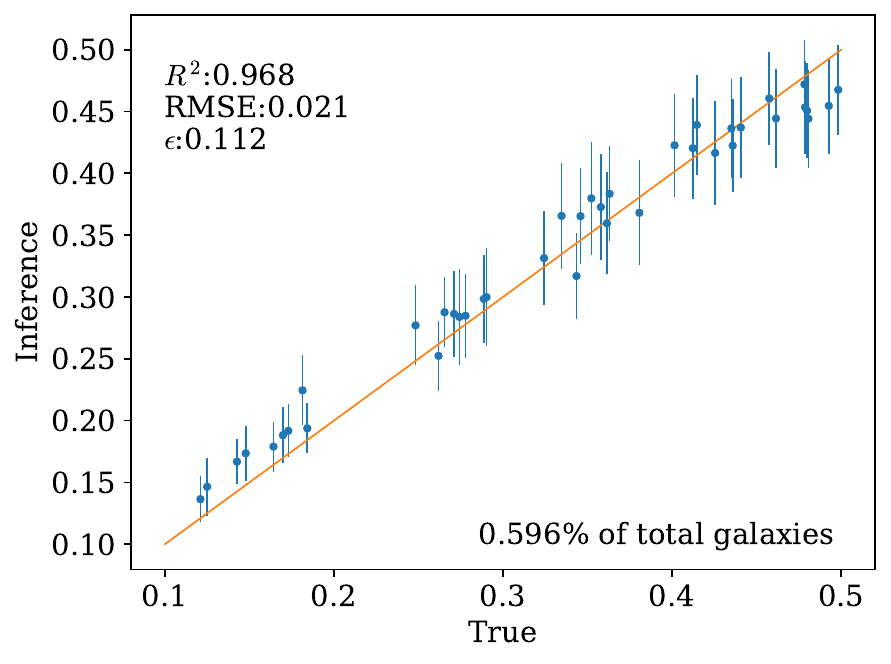}
    \includegraphics[scale= 0.39]{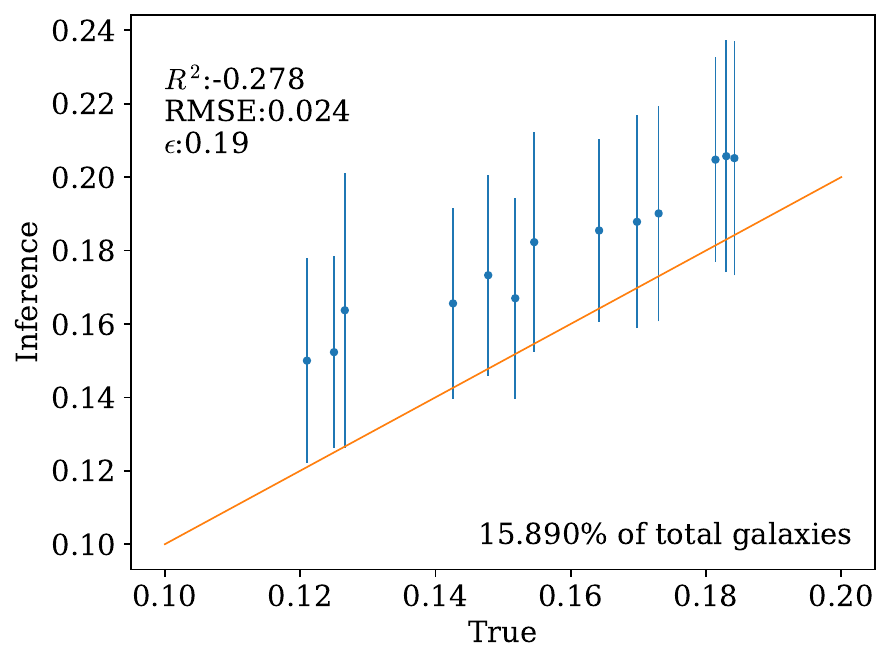}
    \includegraphics[scale= 0.39]{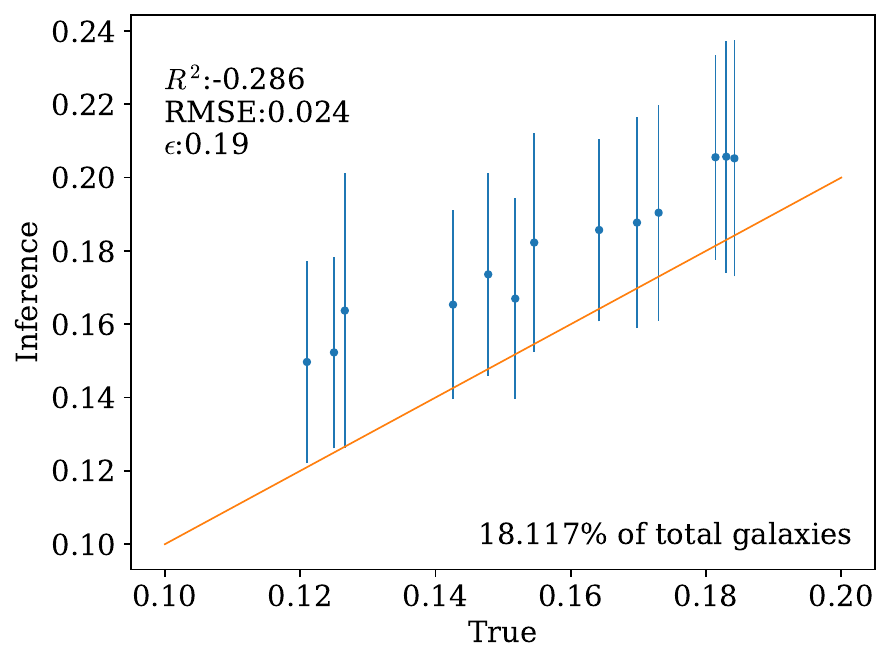}
    \includegraphics[scale= 0.39]{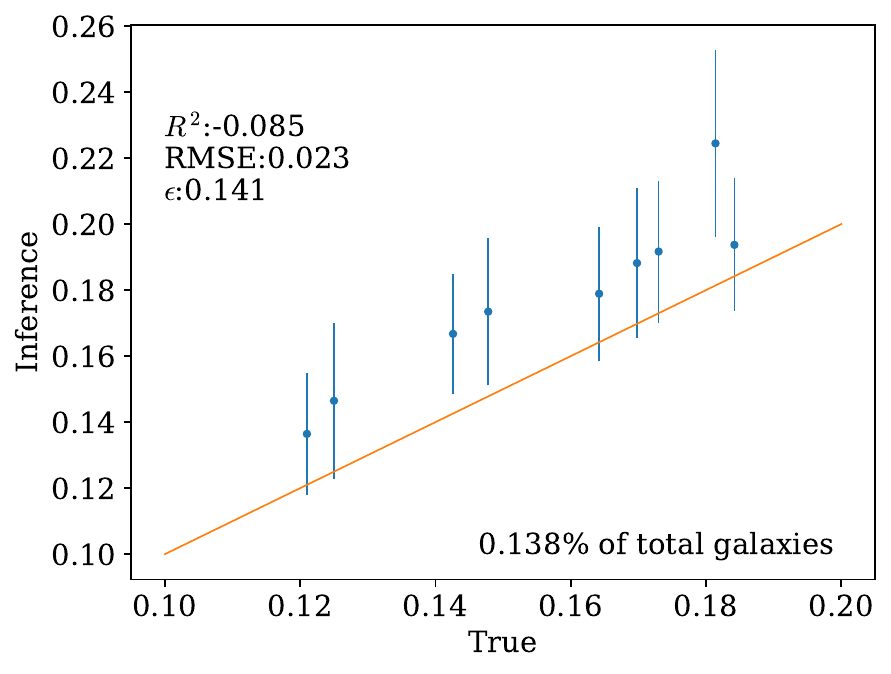}
    
    \caption{The top row shows the average prediction of $\Omega_\mathrm{m}$ using galaxies in the same simulation. The bottom row shows a zoom in on the lower values of $\Omega_\mathrm{m}$. The figure shows an improvement especially on the lower values of $\Omega_\mathrm{m}$ when using void galaxies, that have smaller errorbars (the values of $\epsilon$ highlight the improvement).}
    \label{fig:Mean}
\end{figure*}

In this Section, we use the same trained machine learning model as in \citet{Villaescusa2022}, which takes galaxy properties as inputs and predicts the value of $\Omega_\mathrm{m}$ with fully connected layers. As described in Section \ref{sec:Void Galaxy Selection}, we have classified galaxies into three groups: void, non-void, and random galaxies. Random galaxies refer to galaxies chosen without any specific criteria, irrespective of the environment. We show the prediction of $\Omega_\mathrm{m}$ using these three types of galaxies. To investigate the impact of the galaxy environment on cosmological constraints, we consider galaxies at different levels of isolation. 
This is done by considering two void galaxy selections determined by different cut-off values of the Voronoi cell volume, hence increasing how restrictive the selection is. In particular, we consider as void galaxies the galaxies that have Voronoi cell volumes larger than 100 $(\mathrm{Mpc}/h)^3$ and 150 $(\mathrm{Mpc}/h)^3$ (see Figure \ref{fig:cellDis} for the overall range of volume values), with 150 $(\mathrm{Mpc}/h)^3$ being the most restrictive value, since a higher volume corresponds to more isolated galaxies, and show results for both cases.

Additionally, for the purity cuts at the level of the void catalog we show results after considering only the core density cut (Section \ref{sec:coreDensCut}), or both the core density cut and radius cut (Section \ref{sec:coreDensRadiusCut}). The results show the progressive improvement in constraints on $\Omega_\mathrm{m}$ with those cuts, indicating that a more restrictive void galaxy selection leads to better $\Omega_\mathrm{m}$ predictions. In Section \ref{sec:meanpred}, we show and discuss the average predictions of void, non-void and randomly selected galaxies to investigate the constraining capabilities of the different kinds of galaxies separately.
 
\subsection{Core Density Cut}
\label{sec:coreDensCut}
In this Section, we only apply the core density cut to the void catalog. Figure \ref{fig:CoreDensCut} shows our results. Similar to \cite{Villaescusa2022}, the blue dots and error bars in \Cref{fig:CoreDensCut,fig:coreDensityRadiusCut} represent the predicted posterior mean and standard deviation of $\Omega_\mathrm{m}$ from the networks. The orange diagonal line corresponds to the case in which the predicted values perfectly match the true values. The top left corners of each plot provide the values of accuracy and precision metrics. On the bottom right corners, we indicate the percentages of galaxies considered in the plots. Noticeably, for visualization purposes, in the case of non-void and randomly selected galaxies, we sub-sample the number of points to the same number of data points shown in the plot for void galaxies, allowing a better visual comparison. The metrics are, however, calculated over all the galaxies within each category. 
The left, middle, and right columns show the results of using non-void galaxies, randomly selected galaxies and void galaxies respectively. Therefore, moving from left to right shows the change when considering the \textit{large-scale environment}. 
The first row in Figure \ref{fig:CoreDensCut} shows the results when using different types of galaxies to predict the values of $\Omega_\mathrm{m}$ with a less constrained (smaller) cell volume. From left to right in the row we can see that the result of using void galaxies is generally closer to the reference diagonal line, indicating that using void galaxies we can constrain $\Omega_\mathrm{m}$ better than when using non-void galaxies and randomly selected galaxies. The $R^2$ and RMSE values improve $\sim 5\%$ and $\sim 14\%$ respectively when using void galaxies.

The second row, right column, shows the results when being more restrictive in terms of \textit{local environment}, that is after selecting void galaxies based on a more stringent Voronoi cell volume cut (larger cut, since a larger Voronoi cell volume corresponds to a more underdense environment). 
The improvements seen in the first row from left to right are even more evident in the right plot of the second row---confirming that more isolated void galaxies provide better constraints.

This is consistent with the idea that, to guarantee an isolated environment, void galaxies need to be identified accounting for both the large-scale environment \textcolor{white}{(accounted for when going from left to right in Figure 2 and the local environment (introduced when going from)} (accounted for when going from left to right in Figure \ref{fig:CoreDensCut}) and the local environment (introduced when going from top to bottom in the right column of Figure \ref{fig:CoreDensCut}). We note that in the bottom row, the data points in the figure might be visually less representative for the non-void and random galaxies cases, since for representation purposes we have sub-sampled the points in those cases to show the same number of points as for the void galaxies' case. We also notice that the value of $\epsilon$ is not enhanced for void galaxies compared to randomly selected galaxies. This could be due to the smaller error bars and much lower number of data points in the case of more isolated galaxies (bottom row). The next Section shows results when considering both criteria to increase purity (core density cut and radius cut).  

\subsection{Core Density Cut and Radius Cut}

To further reduce the number of spurious voids in the void catalog, in this Section, we remove voids also based on the radius cut (besides the core density cut). Similar to Section \ref{sec:coreDensCut}, Figure \ref{fig:coreDensityRadiusCut} shows the results of non-void galaxies, random galaxies and void galaxies to infer the values of $\Omega_\mathrm{m}$. The values of $R^2$ and RMSE have a stronger improvement ($R^2 \sim 8\%$ and RMSE $\sim 40\%$ for comparing random galaxies and void galaxies in the bottom row) after this additional purity cut on void radius is applied, showing an even higher constraining power of void galaxies. In addition, in this case, we see that the $\epsilon$ value also improves when considering more isolated galaxies (bottom row, right plot). Therefore, when both criteria for purity are considered (core density cut and void radius), $\epsilon$ improves both when we select void galaxies instead of non-void or randomly selected galaxies (moving from the left and center panels to the right panels) and when we increase the Voronoi cell cut-off volume (from up to bottom panel for void galaxies).
By comparing Figure \ref{fig:CoreDensCut} and Figure \ref{fig:coreDensityRadiusCut}, we conjecture that, since the radius cut removes outlier galaxies with large error bars, it may be responsible for this improvement. If this is the case, the smaller error bars would be the reason for a smaller $\epsilon$ value. As in the case with only the core density cut, the strongest improvement in constraints is obtained when accounting for selections on both the large-scale and local environments (right column, bottom plot).
Therefore, the results of this Section confirm that when using a galaxy from a cosmic void we obtain more stringent constraints.
\label{sec:coreDensRadiusCut}

\subsection{Test on the mean values of predictions with galaxies from the same simulation \label{sec:meanpred}}

In \citet{Villaescusa2022}, the authors show the average prediction of galaxies in the same simulation. Here, to investigate the constraining capabilities on average of the different kinds of galaxies separately (but irrespective of variations of the model of the simulation), we show the average predictions of void, non-void and randomly selected galaxies from the same simulation, that is a simulation where the true value of $\Omega_\mathrm{m}$ is the same, as well as other parameters (see the top row in Figure \ref{fig:Mean}). We note that we perform this binning only for the case in which both purity cuts are applied (radius and core density). In this case, we see a milder improvement when using void galaxies, with the value of $\epsilon$ being slightly boosted. If, however, we zoom in to the lower value of $\Omega_\mathrm{m}$ as shown in the second row of Figure \ref{fig:Mean}, we notice that the error bars on the lower values of $\Omega_\mathrm{m}$ are much smaller when considering void galaxies (seeing an improvement in the bottom row from left to right), corresponding to a further improvement of the $\epsilon$ values. This shows that void galaxies seem to particularly provide better constraints for the lower values of $\Omega_\mathrm{m}$, roughly between 0.1 and 0.2. Yet, we notice that there are some biases for the lower values of $\Omega_\mathrm{m}$, as well as closer to the higher extreme values of $\Omega_\mathrm{m}$. 

\begin{figure}
    \centering
    \includegraphics[scale = 0.45]{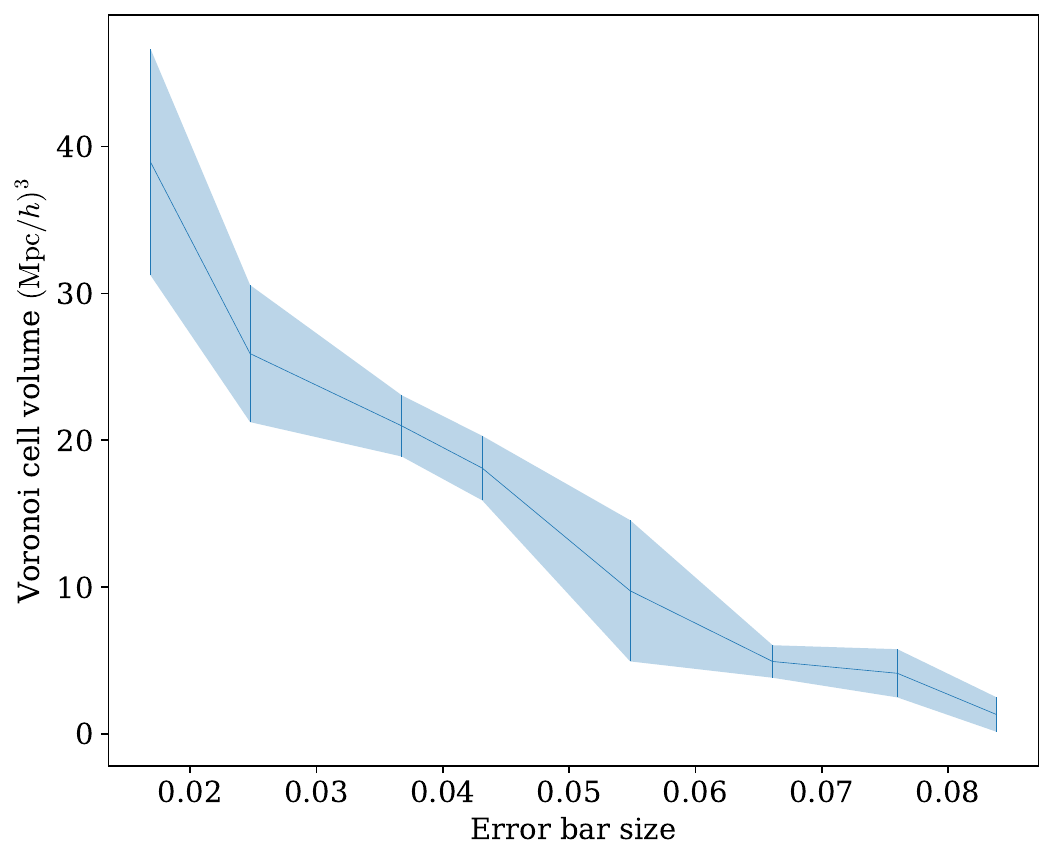}
    \caption{The relationship between the size of error bars from the predictions and the Voronoi cell volume of the galaxies inside voids. We bin the galaxies in each range of error bars and calculate the average Voronoi cell volume of the galaxies in each bin. The error bars are calculated as the standard errors, which consider both the standard deviation and the sample size. The slope of the curve shows that the error bars of the predictions (on the x-axis) are smaller for galaxies with larger Voronoi cell volume (on the y-axis), that is showing that more isolated void galaxies provide more stringent constraints.}
    \label{fig:CellError}
\end{figure}

Furthermore, in Figure \ref{fig:CellError} we plot the error bars of the predictions as a function of the Voronoi cell volume of the galaxies (considering galaxies within 75\% of the void radius from void centers as above). A correlation is present: we notice that the error bars of the predictions (x-axis) are smaller for galaxies with larger Voronoi cell volume (y-axis)---indicating that more isolated void galaxies are more constraining. The scale at which a void galaxy is isolated (from a local environment perspective) corresponds to Voronoi cell volumes of 40 $(\mathrm{Mpc}/h)^3$, that is to a scale of $\sim 5 \mathrm{Mpc}$, considering $h \approx 0.7$. This provides additional evidence that galaxies in underdense regions are better for predicting $\Omega_\mathrm{m}$. 

\section{Conclusion}\label{sec:conc} 
In this work, we have shown that the correlation between $\Omega_{\rm m}$ and galaxy properties exhibit less scatter for void galaxies than for non-voids galaxies. This is an important and new result that may shed light on the impact of baryonic processes on galaxy formation and evolution as a function of the cosmic environment. Recent works showing that void galaxies have distinct characteristics in the IllustrisTNG simulation \citep{Rodriguez-Medrano2023, Curtis2024} could hint to an explanation for our results. We also show that the improvement obtained when using void galaxies is more significant when a more stringent selection is considered, robustly selecting according to both the large-scale and local environments. This improvement, seen when more conservatively selecting void galaxies, seems to confirm the physical expectation that void galaxies provide a cleaner environment to extract cosmological information. Additionally, we notice that the constraining power of void galaxies seems higher in the case of lower values of $\Omega_\mathrm{m}$, a result that requires further investigation and paves the way to future studies on the topic. 

Our results imply that, if we consider void galaxies, we can obtain similar accuracy and precision for cosmological inference as when using non-void or randomly selected galaxies, but with much less data. It also shows that, for an equal number of galaxies selected, the constraining power obtained by galaxies in voids is stronger than the one obtained when using random galaxies. The implications of our results therefore favor the interpretation that void galaxies provide a cleaner and more effective environment for extracting
cosmological information. These results should be further tested on different simulations, such as {\tt\string ASTRID} and {\tt\string SIMBA}. 

Looking ahead to applications to data, it would be interesting to investigate whether these results are expected to hold with observational data from e.g. photometric surveys \cite[see][]{Hahn2023}. Our results can serve as an indication to observers wishing to robustly select void galaxies by taking into account both the large-scale void structure and the local environment. Additionally, we notice that, with respect to observation strategies, there is a trade-off between using all available galaxies versus restricting to void galaxies: while void galaxies yield better constraints, they are scarcer. For more stringent constraints, the strategy could directly aim at targeting void galaxies. Finally, our results may, in the longer term, motivate observation campaigns such as CAVITY \citep{Perez2024}, DIVE \citep{DeLosReyes2023} and Nearby Voids \citep{Pustilnik2019} to observe sky patches in underdense regions, but specifically designed to extract cosmological constraints.

\section{Acknowledgments}\label{sec:ack}
We are especially grateful to  Francisco Villaescusa-Navarro, whose invaluable help and support were pivotal to the completion of this paper. We also thank Wenhan Zhou, Tri Nguyen, Eiichiro Komatsu, Leander Thiele, ChangHoon Hahn, and Sohaib Bhatti for valuable discussions. We are grateful to the anonymous referee for their constructive report that considerably improved this paper. BW and AP acknowledge support from the Simons Foundation to the Center for Computational Astrophysics at the Flatiron Institute. AP acknowledges support from the European Research Council (ERC) under the European Union's Horizon programme (COSMOBEST ERC funded project, grant agreement 101078174). 

\FloatBarrier
\bibliography{main}{}
\bibliographystyle{aasjournal}
\end{CJK*}
\end{document}